\input harvmac
\newcount\figno
\figno=0
\def\fig#1#2#3{
\par\begingroup\parindent=0pt\leftskip=1cm\rightskip=1cm\parindent=0pt
\baselineskip=11pt
\global\advance\figno by 1
\midinsert
\epsfxsize=#3
\centerline{\epsfbox{#2}}
\vskip 12pt
{\bf Fig. \the\figno:} #1\par
\endinsert\endgroup\par
}
\def\figlabel#1{\xdef#1{\the\figno}}
\def\encadremath#1{\vbox{\hrule\hbox{\vrule\kern8pt\vbox{\kern8pt
\hbox{$\displaystyle #1$}\kern8pt}
\kern8pt\vrule}\hrule}}

\overfullrule=0pt

%
\def\tilde{\widetilde}
\def\bar{\overline}
\def\Z{{\bf Z}}
\def\T{{\bf T}}
\def\S{{\bf S}}
\def\R{{\bf R}}

\font\zfont = cmss10 
\font\litfont = cmr6

\def\bigone{\hbox{1\kern -.23em {\rm l}}}
\def\ZZ{\hbox{\zfont Z\kern-.4emZ}}
\def\half{{\litfont {1 \over 2}}}

\def\C{{\bf C}}

\Title{hep-th/9603150, IASSNS-HEP-96-26}
{\vbox{\centerline{PHASE TRANSITIONS}
\bigskip\centerline{ IN   $M$-THEORY AND $F$-THEORY  
}}}
\smallskip
\centerline{Edward Witten\foot{Research supported in part
by NSF  Grant  PHY-9513835.}}
\smallskip
\centerline{\it School of Natural Sciences, Institute for Advanced Study}
\centerline{\it Olden Lane, Princeton, NJ 08540, USA}\bigskip

\medskip

\noindent

Phase transitions are studied in $M$-theory and $F$-theory.
In $M$-theory compactification to five dimensions on a Calabi-Yau,
there are topology-changing transitions similar to those seen
in conformal field theory, but the non-geometrical phases known
in conformal field theory are absent.  At boundaries of moduli
space where such phases might have been expected, the moduli
space ends, by a conventional or unconventional physical mechanism.
The unconventional mechanisms, which roughly involve the appearance
of tensionless strings, can sometimes be better understood in $F$-theory.

\Date{March, 1996}

\nref\ewitten{E. Witten, ``Some Comments on String Dynamics,''
hepth/9507121.}
\nref\otherstrom{A. Strominger, ``Open $p$-Branes,'' hepth/9512059.}
\nref\hanany{O. Ganor and A. Hanany, ``Small Instantons And
Tensionless Strings,'' hepth/9602120.}
\nref\seibergwitten{N. Seiberg and E. Witten,
``Comments On String Dynamics In Six Dimensions,'' hepth/9603003.}
\nref\newduff{M. Duff, H. Lu, and C. N. Pope, ``Heterotic Phase
Transitions And Singularities Of The Gauge Dyonic String,'' 
hepth/9603037.}
\nref\aspinwall{P. Aspinwall, D. Morrison, and B. Greene ,
``Calabi-Yau Moduli Space, Mirror Manifolds, And Space-Time
Topology Change In String Theory,'' Nucl. Phys. {\bf B416} 
(1994) 414.}
\nref\witten{E. Witten, ``Phases Of $N=2$ Theories In Two
Dimensions,'' Nucl. Phys. {\bf B403} (1993) 159,}
\newsec{Introduction}

In recent investigations of non-perturbative behavior of string theory,
many surprising phenomena have been found. Such phenomena can
very crudely be separated into two kinds.  First are
cases in which the phenomenon itself (such as occurrence of enhanced
gauge theory or extra massless particles
at a particular value of a scalar field) is not
surprising, but its occurrence in a particular string theory,
under particular conditions, came as a surprise.  Extended gauge
symmetry or extra massless matter resulting from a Calabi-Yau singularity
or a small heterotic string instanton or strong heterotic string
coupling are examples of this kind.  
Associated with all these surprises is generally a meta-surprise which
is simply our ability to understand the phenomenon!

The second kind of surprises are those  in which
the phenomenon that is found to occur in string theory was not previously
known to be possible in any physics model under any conditions.
These include critical points and phase transitions that cannot
be described by weakly coupled field theory of any sort,
roughly because the objects that are becoming light are strings
instead of particles
\refs{\ewitten - \newduff}.  For example, in six dimensions,
practically any dynamics involving tensor multiplets is exotic
in this sense, and much can be learned from rather general
arguments \seibergwitten.

The main goal of the present paper is a more microscopic study of some
such phenomena in five and six dimensions, using $M$-theory and $F$-theory.
In the $M$-theory case, we begin by studying
in general  the vector moduli space of $M$-theory
compactified to five dimensions on a Calabi-Yau manifold $X$.
Because of the usual relation of $M$-theory to Type IIA, this compactification 
is a certain strong coupling limit of the Type IIA superstring
compactified to {\it four} dimensions on $X$.
The vector moduli space of Type IIA superstring theory 
is extremely rich \refs{\aspinwall,\witten}, with abundant ``phase
transitions'' between a variety of geometrical and non-geometrical
phases (that is, phases describable as sigma models with Calabi-Yau
target and phases that must be described as more abstract conformal
field theories).  We put the phrase ``phase transitions'' in quotes
because in the four-dimensional context, these are not true phase
transitions, though they look like sharp phase transitions in a mean field 
theory approximation \witten.

As we will see in section two, 
in $M$-theory compactified to five dimensions on $X$,
there is a somewhat similar story, in developing which we draw on and
extend certain results obtained recently 
\nref\fone{
     A.C. Cadavid , A. Ceresole , R. D'Auria , S. Ferrara,
``Eleven-Dimensional Supergravity Compactified on Calabi-Yau Threefold,''
hepth/9506144.}
\nref\ftwo{I. Antoniadia, S. Ferrara, and T. R. Taylor,
  ``$N=2$ Heterotic Superstring and its Dual Theory in Five Dimensions,''
hepth/9511108.}
\nref\fthree{S. Ferrara, R. R. Khuri, and R. Minasian,
   ``$M$-Theory on a Calabi-Yau Manifold,'' hepth/9602102.}
\refs{\fone - \fthree}.
The five-dimensional case, however,  has some crucial novelties.
First of all, in the five-dimensional case, one does get sharp
phase transitions between the different
geometrical phases.  Second (as one might suspect from the fact that
abstract conformal field theory has no evident role in $M$-theory), the 
non-geometrical phases do not appear in the five-dimensional story.
Third and  especially, at the overall boundaries
of the moduli space (beyond which in four dimensions the non-geometrical
phases appear), one gets a ``surprise of the second kind,'' 
 outside the scope of conventional low energy effective field theory.
This involves   a critical point at which an infinite number of
particles of arbitrarily high spin go to zero mass, including
electric charges and states of a dual magnetic string.

The earlier  examples of ``phenomena of the second kind''
\refs{\ewitten - \newduff} have generally been in six dimensions and
involve light strings that have both electric and magnetic couplings.
Six dimensions is a very natural dimension for non-critical strings,
because the interesting six-dimensional
non-critical strings (whose tension vanishes
at some point in moduli space) couple to a tensor multiplet, which
is special to six dimensions.  One may therefore
wonder if the tensionless strings seen in $M$-theory could be better
understood by lifting the picture to six dimensions (just as one
has at this point already lifted the traditional problem of the vector
moduli space from four dimensions to five in going from Type IIA to
$M$-theory).  Fortuitously, a remarkable construction known
as $F$-theory
\ref\vafa{C. Vafa, ``Evidence For $F$-Theory,'' hepth/9602022.}
does make it possible to lift the whole discussion to six dimensions,
at least for a suitable class of Calabi-Yau manifolds.

In \ref\morrisonvafa{D. 
Morrison and C. Vafa, ``Compactification
Of $F$-Theory On Calabi-Yau 
Threefolds, I,'' hepth/9602114.}, the heterotic
string compactified to six dimensions on K3 was related to $F$-theory
compactified to six dimensions on certain Calabi-Yau manifolds $X$.
This is a very interesting case for the study of non-critical
strings, because the heterotic string on K3 has (for most
values of the instanton numbers) a strong coupling singularity
\ref\minasian{M. Duff, R. Minasian, and E. Witten, ``Evidence For
Heterotic/Heterotic Duality, hepth/9601036.}
at which apparently
\refs{\seibergwitten,\newduff} a string goes to zero tension.  
It also apparently has a transition in which an instanton turns
into an $M$-theory five-brane, again via appearance of a tensionless
string \refs{\hanany,\seibergwitten}.
The general arguments used previously
do not give very precise information about what kind of string
goes to zero tension at these singularities, 
but such information can be extracted from $F$-theory.  We will
see, for instance, that in certain cases the non-critical
strings that go to zero tension at the strong coupling singularity
are objects that have been encountered before, and in other cases
they are new.

In section four, we go on to consider the occurrence of non-critical
strings of vanishing tension in string compactification to four dimensions.
One easy example (in view of \ref\becker{K. Becker, M. Becker,
and A. Strominger, ``Fivebranes, Membranes, And Nonperturbative
String Theory,'' hepth/9507158.} and \ewitten) is
the Type IIB superstring on a Calabi-Yau manifold, which develops
a tensionless non-critical string when one approaches a conifold
singularity from the Kahler side.  The same tensionless
string arises in heterotic string compactification on a certain
Calabi-Yau manifold, as we show by considering an $F$-theory dual.

\newsec{The Vector Moduli Space Of $M$-Theory}

\subsec{Generalities}

\def\P{{\bf P}}
The Kahler metric of a Calabi-Yau manifold $X$ (for given complex
structure) depends on $b_2={\rm dim}\,H^2(X)$ 
parameters, which determine the
cohomology class of the Kahler form in the cohomology group 
$H^2(X,\R)$ .  One function of these parameters is the 
volume of $X$, and is associated with a hypermultiplet, while the
other $b_2-1$ parameters, which control the ``shape'' of $X$,
are associated with the vector multiplets \fone.  To study
the vector moduli space, 
we are thus mainly interested in varying the shape, without letting
the overall volume go to zero or infinity.   

In Calabi-Yau compactification of
Type IIA superstring theory to four dimensions, there are theta
angles that are related by supersymmetry to the Kahler class of the
metric; upon including them, one describes the vector moduli space
in terms of the complexification of $H^2(X,\R)$.  In Calabi-Yau
compactification of $M$-theory to five dimensions,
the theta angles are absent, so the real parameters of
$H^2(X,\R)$ (with the volume scaled out) are the relevant ones.

For a given Calabi-Yau manifold $X$, the possible Kahler metrics
fill out a ``cone'' in  $H^2(X,\R)$.  As one approaches the boundary
of the cone, $X$ develops a singularity.  In Type IIA compactification
on $X$, the sigma model of $X$ is singular only when the classical
manifold $X$ is singular and in addition a certain theta angle
vanishes.  By using a generic value of the theta angle, one can
go smoothly past the singularity to get to another ``phase'' of
the conformal field theory.  
This phase is defined in a region that is outside the Kahler cone
of $X$; it might be the Kahler cone of another Calabi-Yau manifold
$Y$ (in which case this transition is a topology-changing process
$X\to Y$), or it might be associated with a more abstract conformal
field theory (such as a Landau-Ginzburg model).

The five-dimensional vector multiplet contains only one scalar;
upon compactification to four dimensions, it gains a second scalar,
related to the world-sheet theta angle of the Type IIA sigma model.
This means that effectively in five dimensions, the theta 
angles are frozen to zero, so there is no way to go around
the singularities.  Any continuation from one phase to another
will necessarily involve going through the singularities.  That
is why in five dimensions we will get sharply defined phases
and phases transitions, which are smoothed out if one compactifies
to four dimensions (roughly as, for instance, a standard ferromagnetic
phase transition in three dimensions in a system with continuous
symmetry is smoothed out if one
compactifies to two dimensions).

There are various possibilities for how $X$ may behave as one
approaches the boundary of the  Kahler cone.  We will in this paper
consider only the case that in the limit $X$ is a complex three-manifold
with singularities (as opposed to the possibility that    $X$ collapses
to a space of complex dimension less than three on the boundary of
the Kahler cone).  The possible singularities can be very crudely
classified as follows:

(1) It may be that a complex curve $E$ is collapsing to a point
as one approaches the boundary of the Kahler cone.

(2) It may be that a complex divisor $D$ is collapsing, either
to (a) a curve, or (b) a point.

Case (1) is the case of   topology change -  on the other side of
the boundary of the Kahler cone, one has the Kahler cone of a
different (but birationally equivalent) Calabi-Yau manifold $Y$.
In section 2.2, we will see how this topology change can be described
in $M$-theory.                 

Let us call the union of the Kahler cones of all Calabi-Yau's
that are birationally equivalent to $X$ the ``extended Kahler cone''
of $X$.  When one approaches the boundary of the extended
Kahler cone, one gets a singularity of type (2).  In Type IIA
compactification on $X$, one can  continue past type (2) singularities,
and then one sees phases based on 
more abstract conformal field theories
rather than sigma models.  We will explain in section 2.3 that
in $M$-theory most or possibly all of the more abstract phases  
are absent.  So 
the vector moduli space of $X$ is just the extended Kahler cone
(possibly with a few but not all of the non-geometrical phases
added).

This means in particular that the vector moduli space is, with
its natural metric, an {\it incomplete} manifold.  One can reach
the boundary in a finite distance.  
What physics can be associated
with this?  One way to get a boundary in the moduli space is
to find an enhanced $SU(2)$ gauge symmetry.   In supersymmetric
$SU(2)$ gauge theory in five dimensions, the only scalar field is
a field $\phi$ in the adjoint representation of $SU(2)$; the moduli
space of classical vacua is thus parametrized by the
order parameter $u=\tr \phi^2$, which is real and non-negative,
so the moduli space is the half-line $u\geq 0$.  An $SU(2)$ gauge
symmetry is restored at the boundary point $u=0$. 
It has indeed been shown that a boundary of the Kahler cone
 of type (2a) is associated with enhanced
$SU(2)$ gauge symmetry \ref\kmp{S. Katz,
D. Morrison, and R. Plesser, ``Enhanced Gauge Symmetry In Type II
String Theory,'' hepth/9601108.}. (See also \ref\aspinwalla{P. 
Aspinwall, ``Enhanced Gauge Symmetry And Calabi-Yau
Three-Folds,'' hepth/9511171.} 
for the case of collapse of a divisor to a curve
of genus zero, and \ref\klma{A. Klemm and P. Mayr,
``Strong Coupling Singularities And Non-Abelian Gauge Symmetries
In $N=2$ String Theory,'' hepth/9601014.} for related issues.)

Note that in five-dimensional supersymmetric
field theory, to restore a gauge symmetry of rank greater
than one requires adjusting more than one real parameter; thus $SU(2)$
is the only extended gauge 
symmetry that can appear on going to the boundary
of the extended Kahler cone in a generic fashion. More generally,
as long as the physics is free in the infrared, and so describable
by an effective classical field theory, restoration of an $SU(2)$ gauge
symmetry or a discrete $\Z_2$ symmetry is the only way to produce
a boundary of the moduli space.  Thus the fact that a singularity
of type (2a) gives precisely an $SU(2)$ gauge symmetry is no accident.

There remains the case (2b), which is possibly more typical, and 
brings us finally to a ``surprise of the second kind'' as 
promised in the introduction.
As we will explain in section 2.4,
when one approaches a singularity of type (2b), one gets a novel
kind of low energy physics with infinitely many particles of
arbitrarily high spin all going to zero mass.  There is also
a tensionless non-critical string which is ``magnetically'' charged
with respect to this infinity of light ``electric'' charges.  The limiting
theory where all these particles reach zero mass is (as in like examples 
mentioned in the introduction) 
not free in the infrared and so  beyond the reach of the comments 
in the last paragraph.

\subsec{Collapse Of A Curve}

In compactification of $M$-theory on a Calabi-Yau manifold
$X$ to a five-manifold $W$, 
vector fields $A^a$, $a=1,\dots, {\rm dim}\,H^2(X)$ arise
from the five-dimensional reduction of the underlying three-form
field $C$.  These vector fields interact, among other things,
through Chern-Simons couplings
\eqn\normo{L_{CS}= {1\over 24\pi^2}\int_Wd^5x\, 
\epsilon^{\mu\alpha\beta\gamma
\delta}A_\mu^a\partial_\alpha A_\beta^b\partial_\gamma A_\delta^c
\lambda_{abc}.}
Here the $\lambda_{abc}$ are constants (integers as discussed later)
 determined
by the intersection ring of $X$; they in turn determine the
metric on the vector moduli space.  

Now consider the behavior as one approaches the boundary of the Kahler
cone and a complex curve $E$ collapses.  A BPS-saturated
hypermultiplet, which arises by wrapping a two-brane over $E$, goes
to zero mass in this limit.  The effective Lagrangian of a 
five-dimensional hypermultiplet $H$ can depend on a real  mass parameter $m$,
which enters by terms $m^2|\phi|^2+m\bar\psi \psi$, where $\phi$
and $\psi$ are the bosons and fermions in $H$.  The particles
obtained by quantizing $\phi$ and $\psi$
have mass $|m|$.  Thus, continuation to negative $m$ 
makes sense in the low energy effective action.  

What would a continuation to negative $m$ mean in terms of $M$-theory,
in the present context?  For the
hypermultiplet $H$ that arises by wrapping a two-brane over $E$, $m$
is the area of $E$.  Thus the continuation to negative $m$ is a kind
of continuation to negative area.  This has been
encountered in studying the vector moduli space in Type IIA superstring
theory: the continuation to negative area is a ``flop'' to a different
(but birationally equivalent) Calabi-Yau manifold $Y$. In 
this transition, the curve $E$, whose area should
superficially become negative, disappears and is replaced by a curve
$F$ on $Y$ of positive area but opposite cohomology class to $E$.  
\foot{
An example of this phenomenon is given at the end of the present paper.}

To clarify the physical meaning of the sign of $m$, recall that in five 
dimensions the Lorentz group $SO(1,4)$ has only one 
spinor representation, which is pseudo-real.  The Clifford
algebra $\{\Gamma_\mu,\Gamma_\nu\}=2\eta_{\mu\nu}$, has, however,
two representations, one with $\Gamma_0\Gamma_1\Gamma_2\Gamma_3\Gamma_4
=1$, and the other with $\Gamma_0\Gamma_1\Gamma_2\Gamma_3\Gamma_4
=-1$.  The two representations differ by $\Gamma_\mu\to -\Gamma_\mu$,
which preserves the anticommutation relations.  
While $SO(1,4)$ has only one spinor representation, $SO(4)$, which
is the little group of a massive particle, has two.  Once a representation
of the Clifford algebra is picked, the Dirac equation with mass reads
$\left(i\Gamma^\mu D_\mu-m\right)\psi=0$, 
and the sign of $m$ determines under
which of the two spinor representations of $SO(4)$ the massive particle
transforms.  Thus, the two signs of $m$ are physically inequivalent,
but which sign of $m$ goes with which representation of $SO(4)$ depends
on which representation of the Clifford algebra one uses -- since
$\Gamma^\mu\to -\Gamma^\mu$ has the same effect on the Dirac equation
as $m\to -m$. 

Now, the manifolds $X$ and  $Y$ -- with $Y$ being obtained from $X$
by ``analytic continuation to negative area'' -- have different
intersection forms, and therefore different values of the
coefficients $\lambda_{abc}$ of equation \normo.  If  the 
$M$-theory on $X$ looks at low energies like a theory with
a hypermultiplet of mass $m$ interacting with gauge fields, and if
continuation past $m=0$ corresponds to jumping
from $X$ to $Y$, then we must see that coupling of gauge fields to
a charged hypermultiplet results in a jump in the Chern-Simons
coefficients when one passes through $m=0$.  

This is implicit in computations in \ftwo\ and is not
hard to verify  directly.  Suppose that our hypermultiplet
$H$ couples to a linear combination $A=\sum_a c_aA^a$ of the vectors,
with the $c_a$ being coefficients.  Consider the one-loop
$AAA$ amplitude due to the charged fermion loop.  
\foot{On dimensional grounds in this unrenormalizable theory,
or by arguments involving locality or the quantization of the
Chern-Simons coupling, diagrams with more loops cannot renormalize
the Chern-Simons interaction.}
If the three
external photons have momenta $p,q$, and $-p-q$, and polarizations
$\alpha,$ $\beta$, and $\gamma$, then the diagram in which they are
attached to the fermion loop in that cyclic order gives
an amplitude
\eqn\nitro{{1\over (2\pi)^5}\int d^5k\,\tr\Gamma_\mu {1\over \Gamma\cdot
(k+p)-m}\Gamma_\alpha {1\over \Gamma\cdot k-m}\Gamma_\beta{1\over
\Gamma\cdot (k-q)-m}.}
After doing the Dirac algebra, the parity-violating part of this is
\eqn\hitro{-{m\epsilon_{\mu\alpha\beta\gamma\delta}p^\gamma q^\delta
\over 8\pi^5} \int {d^5k\over (k^2-m^2)^3}}
plus terms of higher order in external momenta.
Notice that an explicit factor of $m$ appears in the numerator;
on the other hand, after Wick rotation, the integral in \hitro\
can be evaluated to give $\pi^3/2|m|$.  Thus, the amplitude
is proportional to $ m/|m|={\rm sign}\,m$ and is in fact
\eqn\jitro{-{i\,{\rm sign}\, m\over 16\pi^2}\epsilon_{\mu\alpha\beta
\gamma\delta}p^\gamma q^\delta.}  After adding the crossed diagram,
this corresponds to an $A^3$ interaction vertex
\eqn\pitro{-{{\rm sign}\,m\over 48\pi^2}\int_Wd^5x\,\, \epsilon^{\mu
\alpha\beta\gamma\delta}A_\mu\partial_\alpha A_\beta \partial_\gamma
A_\delta.}  The jump when $m$ changes sign is thus
\eqn\juryduty{{1\over 24\pi^2}\int_W d^5x \,\,\epsilon^{\mu
\alpha\beta\gamma\delta}A_\mu\partial_\alpha A_\beta \partial_\gamma
A_\delta.} 

This should be compared with the change in the intersection form
under birational transformation from $X$ to $Y$.  The change is that,
for every curve $E$ that collapses, the Yukawa coupling of the
multiplet containing the corresponding $A$ changes by 1. (See,
for example, pp. 209-211 of \witten.)
This agrees with \juryduty\ provided that the expression \juryduty\
is the correctly normalized five-dimensional Chern-Simons action
with coefficient 1.  In verifying this, there are some subtleties.
Recall that the Chern-Simons integral on a five-manifold $W$ can be
usefully defined in terms of the integral of $F\wedge F\wedge F$ over
a six-manifold $Z$ with boundary $W$; 
the normalization of the Chern-Simons
action is conventionally
chosen so that this integral is independent
of $Z$ (and of the extension of the gauge field over $Z$) precisely modulo
$2\pi$.  The expression
\eqn\curryduty{{1\over 4\pi^2} \int_Zd^6x\,
\epsilon^{\mu\nu\alpha\beta\gamma
\delta}\partial_\mu A_\nu \partial_\alpha A_\beta \partial_\gamma A_\delta
} is correctly normalized so that,  for  closed   oriented
six-manifolds $Z$ and $U(1)$ gauge fields $A$, its possible values are 
 $2\pi n$
for arbitrary integer $n$.  So, because the denominator is
$24\pi^2$ instead of $4\pi^2$, \juryduty\ is
$1/6$ of the Chern-Simons interaction as it would conventionally be
defined.

This seems to have the following interpretation.  Suppose that the closed
oriented six-manifold $Z$ is actually a spin manifold with $p_1=0$.
\foot{In what follows, we will not analyze the torsion and
so will not get a complete result.  One should really use  the 
characteristic class $p_1/2$ (which is well-defined for spin manifolds) 
rather than $p_1$.  The argument given momentarily
that $c_1(L)^3$ is divisible by six only requires
that $p_1/2$ vanish modulo torsion.  For the cobordism assertion
of the next paragraph, $p_1/2$ should vanish; I do not know if
the statement still holds if $p_1/2$ is a torsion class.
The $M$-theory argument at the end of this sub-section only shows
that $p_1(W)/2$ can be assumed to vanish mod torsion; I do not know if
$M$-theory allows torsion in $p_1(W)/2$.}
Then one can show that \curryduty\ is divisible by $12\pi$, and
not just $2\pi$.  This amounts to the following topological fact.
If $L$ is a complex line bundle over an oriented  
six-manifold $Z$, then in general
$c_1(L)^3$ can be an arbitrary integer; but if $Z$ is spin and has
$p_1=0$, then $c_1(L)^3$ is divisible by six.  Indeed, on a six-manifold
$Z$ that is spin with $p_1=0$,  the index of the Dirac operator, for spinors
with values in $L$, is $c_1(L)^3/6$, showing that $c_1(L)^3$ is
divisible by six.  

Suppose, then, that one wants to define the Chern-Simons interaction
not on arbitrary oriented five-manifolds $W$ but only for those 
that are spin manifolds with $p_1=0$.  If $W$ has 
the stated properties, it can be shown using cobordism theory to be the
boundary of a six-manifold $Z$ that is likewise spin, with $p_1=0$,
and by using only such $Z$'s in defining the Chern-Simons interaction
on $W$, one can ensure that \juryduty\ is uniquely defined
modulo $2\pi$, even though the denominator $24\pi^2=6 \cdot 4\pi^2$ in
\juryduty\ is six times the denominator in \curryduty.

It remains, then, to explain why in $M$-theory on $W\times X$,
with $W$ a five-manifold and $X$ a Calabi-Yau manifold, we are only
interested in the case that $W$ is spin and has $p_1=0$.  $W$ must
of course be spin because the theory has fermions.  The restriction
to $p_1=0$ comes because 
\nref\vafawitten{C. Vafa and E. Witten, ``A One-Loop Test Of
String Duality,'' Nucl.Phys. {\bf B447} (1995) 261.}
\nref\duffetal{M. J. Duff, J. T. Liu, and R. Minasian,
 ``Eleven-Dimensional Origin Of String-String 
Duality: A One Loop Test'' Nucl. Phys. {\bf B452} (1995) 261.}  
\refs{\vafawitten,\duffetal} in $M$-theory there is an interaction
$C\wedge I(R)$ where $C$ is the three-form of $M$-theory and $I(R)$ is
an eight-form constructed as 
a quartic polynomial in the Riemann tensor.  Because 
of this interaction, the equations of motion can only be obeyed if
$I(R)$ vanishes cohomologically.  In compactification on $W\times X$,
$I(R)$ is a multiple of $p_1(W)\cdot p_1(X)$, and as $p_1(X)\not= 0$
for an arbitrary Calabi-Yau manifold $X$, it follows that one must
require $p_1(W)=0$.

\subsec{Absence Of The Non-Geometrical Phases}

In Type IIA compactification on a Calabi-Yau $X$, there are many geometrical
and non-geometrical phases.  We have just seen that the geometrical
phases are connected in $M$-theory (though, unlike the Type IIA case,
one must go through true phase transitions to connect them).  About
the non-geometrical phases, one  faces a puzzle: they are described
in string theory by relatively abstract conformal field theories
(rather than sigma models), and it is hard to see what this could
correspond to in $M$-theory.  We will now argue that the non-geometrical
phases are absent in $M$-theory.  (It follows that the non-geometrical
phases are also absent in $F$-theory, which turns into $M$-theory
upon compactification on a circle.) 

Before getting into a general discussion, let us first mention an
important special case of how to see the absence in $M$-theory
compactification to five dimensions of a continuation to a non-geometrical
phase.  At  certain kinds of boundaries of the
generalized Kahler cone \kmp, one gets an enhanced $SU(2)$ gauge
symmetry.  (A derivation of this result is given in the next subsection.)
As explained in section 2.1, the order parameter for such a symmetry
enhancement is a real, positive semi-definite field $u=\tr \phi^2$,
with $\phi$ a {\it real} scalar in the adjoint representation.  Thus,
one sees in the low energy field theory that $u=0$ is a boundary
of the moduli space, with no way to continue beyond it.  After
compactification to four dimensions, $\phi$ and $u$ become complex
and one can continue past $u=0$ to a non-geometrical phase.

Now let us analyze why such boundaries appear.
In Type IIA superstring theory on $\R^4\times X$,  the number
of vector multiplets is $b_2(X)$, and the vector moduli space is
modeled on $H^2(X,\C)$.  In $M$-theory on $\R^5\times X$, there are
only $b_2(X)-1$ vector multiplets; the overall volume of $X$ transforms
in a hypermultiplet \fone, while the scalars in vector multiplets
are the ``shape'' parameters in $H^2(X,\R)$.  The vector moduli
space is thus roughly the projectivization (in the real sense)
of $H^2(X,\R)$.

In a given geometrical phase, in taking a large volume on $X$,
one can read off from eleven-dimensional supergravity the metric
on the $M$-theory vector moduli space \fone.  The result so obtained
is exact, in that phase, since the unbroken five-dimensional supersymmetry
permits
no corrections.  Indeed, as the volume of $X$ transforms in a hypermultiplet,
the metric on the vector moduli space is independent of the volume,
so can be computed in the large volume, field theory limit.  This
metric is completely determined by the 
Chern-Simons couplings that we discussed in section 2.2 \ref\gunaydin{M.
Gunaydin, G. Sierra, and P. K. Townsend, ``The Geometry
Of $N=2$ Maxwell-Einstein Supergravity And Jordan Algebras,''
Nucl. Phys. {\bf B242} (1984) 244, ``Gauging Of The $D=5$ 
Maxwell-Einstein Supergravity: More On Jordan Algebras,''
Nucl. Phys. {\bf B253} (1985) 573.} 
and thus by the intersection form of $X$.  

In another geometrical phase based on a different birational model
$Y$, the metric on vector moduli space is determined by the intersection
 form
of $Y$.  This is different from that of $X$, so 
the metric and other couplings are non-analytic in crossing
the phase boundary, as befits a {\it bona fide} phase transition.

Now to  compare $M$-theory on $\R^5\times X$ to Type IIA superstring theory
on $\R^4\times X$, we begin by looking at $M$-theory on
$\R^4\times\S^1\times X$.  As the radius, $ R$, of the $\S^1$ goes
to infinity, this goes over to $M$-theory on $\R^5\times X$, while as it goes
to zero, one gets Type IIA on $\R^4\times X$.
If $g_M$ is the $M$-theory metric on $\R^4\times X$, and $g_{II}$ is
the Type IIA metric on $\R^4\times X$, then the relation between them is
(see p. 93 of \ref\dynamics{E. Witten, ``Dynamics Of String Theory
In Various Dimensions,'' Nucl. Phys. {\bf B443} (1995) 85.})
$g_M=g_{II}/T^{1/3}R$.  ($T$ is the two-brane tension, included
here for dimensional reasons.)
If then $K_M$ and $K_{II}$ are the
Kahler classes of $X$ as measured in $M$-theory or in the Type IIA
theory, one has likewise
\eqn\hurry{K_M={1\over T^{1/3}R}K_{II}.}
This means that if one keeps $K_M$ fixed while taking $R$ to infinity,
then $K_{II}$ must go to infinity.  Thus, $M$-theory in five dimensions
only ``sees'' what in conformal field theory would be understood
as the region at infinity in the moduli space.  

In going from the Type II vector moduli 
space to the $M$-theory vector moduli space, 
the overall scale of $K_{II}$
should be eliminated (since the volume is part of a hypermultiplet
in $M$-theory). It is clear from \hurry\ that the scale
of $K_{II}$ is to be removed by  scaling $K_{II}\to\infty$ as $R\to\infty$. 
Note that since the metric on the Type IIA vector moduli space 
depends only on $K_{II}/\alpha'$, we could alternatively take
$\alpha'\to 0$ instead of $K_{II}\to\infty$.

\hurry\ has only been derived so far in a semi-classical sense,
for long wavelengths.  To make an exact statement, one must be more
precise about what $K_{II}$ means.  There are at least two important
sets of natural
coordinate systems on vector moduli space: one can use the coupling
constants of the linear sigma model \witten, which I will call
linear sigma model coordinates but which in 
\ref\measuring{P. S. Aspinwall,  B.  R. Greene, and D. R. Morrison,
``Measuring Small Distances In $N=2$ Sigma Models.''} are called algebraic
coordinates; or one can use the 
special coordinates of $N=2$ special geometry,
which in that paper are called $\sigma$-model coordinates.  Linear
sigma model coordinates  have no natural meaning in $M$-theory, so to make
the most natural comparison between Type IIA and $M$-theory, we should
understand $K_{II}$ to be the Kahler class as defined in special
geometry.  Indeed, the components of $RK_M$ are the special
coordinates of $M$-theory on $\S^1\times X$. 

In linear sigma model coordinates, the geometrical and
non-geometrical phases appear to have a co-equal status:
each occupies a cone in $H^2(X)$.  In special coordinates,
the story is quite different. To within stringy corrections
that move the phase boundaries by an amount of order $\alpha'$,
the geometrical phases occupy in special coordinates the same
cones that they occupy in linear sigma model coordinates, but
the non-geometrical phases are squashed to small regions, with
a thickness of order $\alpha'$, along the boundaries of the 
extended Kahler cone.  This result is derived in \measuring,
and is depicted clearly in figure 6 of that paper.  (For further related
discussion
see \ref\aspin{P. S. Aspinwall, ``Minimum Distances In
Non-Trivial String Target Spaces,'' Nucl. Phys. {\bf B431} (1994) 78.}.)
We will call the squashed cones occupied by the non-geometrical
phases ``plates.''

Now we can easily see what happens in going from Type IIA to $M$-theory.
This is done, as explained above, by scaling $K_{II}$ to infinity,
or equivalently by
taking the limit as $\alpha'\to 0$ with $K_{II}$ fixed.
When we do so, the plates (whose thickness
in at least one direction is of order $\alpha'$, at least in the
cases that have been analyzed) are flattened
to nothing.  
But for $\alpha'\to 0$, the geometrical phases occupy precisely
the cones of the linear sigma model -- the corrections come from
world-sheet instantons whose contributions vanish as $\alpha'\to 0$.

The upshot is that in $M$-theory, the vector
moduli space for any geometrical
phase is simply the corresponding cone in $H^2(X)$ (divided
by overall scaling), with the metric
obtained from eleven-dimensional supergravity. All geometrical
phases must be included, as we saw in section 2.2, and the overall
vector moduli space is the union of all the geometrical phases,
making up the extended Kahler cone.  When one reaches the boundary
of the extended Kahler cone, the moduli space ends, either with
extended $SU(2)$ gauge symmetry \kmp\ or via more exotic
physics that we come to next.

\subsec{Collapse Of A Divisor}

Now we will consider what happens in $M$-theory when a divisor
$D$ collapses as one approaches a boundary of the Kahler cone.
There are two cases to consider: (a) $D$ collapses to a curve $E$;
(b) $D$ collapses to a point.  The former case was studied in
\kmp.

\def\P{{\bf P}}
Simple examples of the two cases are as follows:

(a) $D$ could consist of $\P^1\times F$,  with $F$ a curve.
Then one could consider a limit in which $\P^1$ collapses to 
a point, so that $D$ collapses to $F$.  This can also be generalized
to a fiber bundle, as analyzed in \kmp.  When $D$ collapses in this way,
one gets a curve of $A_1$ singularities (parametrized by $F$).

(b) Consider an isolated $\Z_3$ orbifold point, say the singularity
at the origin in $X'=\C^3/\Z_3$, where $\Z_3$ acts
by $(z_1,z_2,z_3)\to (\omega z_1,\omega z_2,\omega z_3)$, with
$\omega^3=1$.  By blowing up the origin, replacing the origin in $X'$
by a copy of $\P^2$, one gets a smooth (non-compact) Calabi-Yau
manifold $X$, which looks near the $\P^2$ like a piece of a global
Calabi-Yau manifld $X'$.  One can approach a boundary of the Kahler cone
of $X'$ by letting the volume of $D=\P^2$ go to zero, so that $D$ collapses
to a point.  

Now, in either example, let us look for  BPS-saturated states
that can be made by wrapping  two-branes or five-branes on $D$ 
and that go to zero mass as one approaches the singularity.  (See
\fthree\ for some background.)  In case (a), the lightest states come 
from wrapping a two-brane over 
$\P^1\times x$, where $x$ is an arbitrary point in $F$.  The
moduli space ${\cal M}$ of such two-branes
is a copy of $F$.  We will later quantize the collective coordinates
corresponding to this moduli space and recover the spectrum claimed
in \kmp.  If $r$ is the area of the $\P^1$,    such states have
mass proportional to $r$.  One can consider a two-brane wrapped
over a more general complex curve in $D$; this gives a BPS-saturated
state whose mass does not vanish as $ r\to 0$.  Looking for other
light states,
one can make a low-tension string
by wrapping a five-brane over $D$.  Such a string has tension of
order $r$, so (if the usual sort of dimensional analysis can be
applied) states obtained as excitations of such a string have
masses of order $r^{1/2}$.  Thus the lightest states, for $r\to 0$,
come from the two-brane wrapped on a copy of $\P^1$.

Now we come to case (b).  First of all, the opportunities for two-brane
wrapping are much richer.  Let $r$ be the Kahler class of $D=\P^2$,
and let $(y_1,y_2,y_3)$ be homogeneous coordinates for $D$.  Then one
can in a supersymmetric fashion wrap a two-brane over any complex
curve in $D$ given by a degree $n$ homogenous equation $f_n(y_1,y_2,y_3)
=0$.  Let ${\cal M}_n$ be the moduli space of such curves; thus
the dimension of ${\cal M}_n$ is
\eqn\jdin{{\rm dim}\,{\cal M}_n={n(n+3)\over 2}.} A two-brane
wrapped over such a degree $n$ curve has area $nr$, so quantization
of ${\cal M}_n$ will give states whose mass is $nr\cdot T_2$, with
$T_2$ the two-brane tension.  The fundamental difference from the
previous case is that all values of $n$ arise, and not only $n=1$.
\foot{In case (a), one can consider several parallel two-branes,
each wrapped over a copy of $\P^1$, and look for a bound state
-- which would give $n>1$ -- but it is believed that such bound
states do not exist.  In case (b) one gets bound states for free
simply because the generic $f_n$ is irreducible, so that the wrapped
two-brane is not a combination of objects of lower degree.}  In case
(b), since the volume of $D$ is $r^2/2$, one can make a string
with tension of order $r^2$ by wrapping a five-brane over $D$. 
If the usual dimensional analysis holds, states obtained
by quantizing such a string have masses of order $(r^2)^{1/2}=  r$,
just like the two-brane states.  

Because of duality between two-branes
and five-branes (and the fact that the complex curve $f_n=0$
has a non-zero intersection number with  $D=\P^2$),
the string obtained from the five-brane is
``magnetically'' charged with respect to the ``electric'' charges
that come from the two-branes.  Thus, while in case (a) 
one gets eventually  near the boundary of the Kahler cone
only finitely many light states from quantization of
${\cal M}$, in case (b) we will get infinitely many ``electric''
states from quantizing the ${\cal M}_n$'s, together with
all the modes of the ``magnetic'' string.

\bigskip
\noindent{\it Quantum Numbers Of Electric States}

We will now try to learn a little more by analyzing the quantum
numbers of the ``electric'' states, that is the two-brane wrapping
modes.  (We cannot make a similar analysis for ``magnetic''
states, coming from
the light string, since we do not know how to quantize it; as it
is strongly coupled, it is not clear that it can be quantized
in the conventional sense.)

In Calabi-Yau compactification of $M$-theory, there are eight unbroken
real supersymmetries, transforming as a spinor of $SO(1,4)$.  Under
the little group $SO(4)=SU(2)_1\times SU(2)_2$ of a massless  particle,
the supercharges transform as $2(1/2,0)\oplus 2(0,1/2)$, that is two
copies each of $(1/2,0)$ and $(0,1/2)$.  The presence of a two-brane
breaks half of the supersymmetries, in an $SO(4)$ invariant fashion;
with a suitable choice of orientation we can suppose that the $2(0,1/2)$
supercharges are broken and the $2(1/2,0)$ supercharges annihilate
the classical two-brane configuration.

The breaking of the $2(0,1/2)$ supercharges gives four fermion zero
modes (related by the unbroken supersymmetries to spatial translations),
whose quantization gives four states that transform as $2(0,0)\oplus 
(0,1/2)$, which is the content of a ``half-hypermultiplet'' $H_0$.  The 
rest of the story involves quantization of the collective coordinates of the
two-branes, that is, quantization of the moduli spaces ${\cal M}$
(or ${\cal M}_n$) of complex curves.

There are four unbroken supersymmetries, and they must be realized
in the quantum mechanics of the collective coordinates.  This occurs
in a fairly standard way.  The quantum states are differential forms
on ${\cal M}$.  Because ${\cal M}$ is a Kahler manifold, on the
differential forms there act four natural operators $Q_i=\partial,$
$\bar\partial$, $\partial^*$, and $\bar\partial^*$, which generate
a $0+1$-dimensional supersymmetry algebra and represent the unbroken
$2(1/2,0)$
supercharges in acting on the differential forms on ${\cal M}$.
The BPS-saturated states are the states annihilated by the $Q_i$, that is,
the harmonic forms on ${\cal M}$.

Now we would like to know the spins of these BPS-saturated states.
In the rotation group $SO(4)=SU(2)_1\times SU(2)_2$, both $SU(2)_1$ and
$SU(2)_2$ act trivially on the bosonic collective coordinates of the
two-branes -- the modes tangent to ${\cal M}$.  The story is different
for the fermionic collective coordinates.  As they are generated
from the bosonic ones by the $Q_i$, which transform as $2(1/2,0)$,
the fermionic collective coordinates transform trivially under
$SU(2)_2$, which therefore acts trivially on the harmonic forms on
${\cal M}$.  But by the same token, the fermionic collective coordinates
transform non-trivially under $SU(2)_1$, which therefore acts
non-trivially  on the harmonic forms.

In fact, the action of $SU(2)_1$ on the harmonic forms on ${\cal M}$ is
just the standard action of $SU(2)$ on the cohomology of a Kahler manifold.
One can pick a standard basis $J_3, $ $J_+$, $J_-$ of the Lie algebra
of $SU(2)_1$ so that $J_3$ acts on a $(p,q)$ form on ${\cal M}$
by multiplication by 
$\left((p+q)-{\rm dim}_\C\,{\cal M}\right)/2$, while $J_+$
and $J_-$ act by wedge product or contraction with the Kahler form.
In particular, the $SU(2)_1$ multiplet of highest spin (detected
by the largest $J_3$ eigenvalue) consists of the powers of the Kahler
form and has spin $\half {\rm dim}_\C\,{\cal M}$.  

Now we can analyze the examples (a) and (b) above.
In example (a), with a divisor collapsing to a curve, the relevant
moduli space is 
${\cal M}={ F}$, which is  a Riemann surface of genus, say,
$g$.  The cohomology of ${\cal M}$ 
consists of a $(0,0)$-form, of $ J_3=-1/2$,
a $(1,1)$-form, of $J_3=1/2$, and $2g$ additional 
$(1,0)$ or $(0,1)$-forms,  
of $J_3=0$.  The combined spectrum from quantization
of ${\cal M}$ is thus $(1/2,0)\oplus 2g(0,0)$.
When one tensors this with the half-hypermultiplet  $H_0$ related
to the translations, one gets a vector multiplet, consisting of
 $(1/2,1/2)\oplus
2(1/2,0)$,  and $g$ hypermultiplets, that is, $g$ copies of 
$4(0,0)\oplus 2(0,1/2)$, This is the spectrum obtained in
\kmp, by a different but not totally unrelated method.   The one charged
vector multiplet just found, together with its partner of opposite
charge, generates an enhanced $SU(2)$ gauge symmetry.  The
$g$ hypermultiplets, together with their charge conjugates and
some neutral modes that come from complex structure deformations, 
make $g$ hypermultiplets
in the adjoint representation of $SU(2)$.

In example (b), with a divisor collapsing to a point,
we see using \jdin\ that the maximum value of $J_3$ in quantizing
${\cal M}_n$ is $n(n+3)/4$.  Since the mass is $M\sim n$,
this gives the relation -- familiar from quantization of strings
-- $J\sim M^2$ for large $M$.  

Along with the ``electric'' states that we have just analyzed,
the same model also has ``magnetic'' states associated with the light
string.  How could one hope to find a common origin for the electric
and magnetic states together?
The only obvious hope is to use $F$-theory, replacing $M$-theory
on $\R^5\times X$ with $F$-theory on $\R^5\times \S^1\times X$.
Here, as explained in \vafa, $X$ must be a Calabi-Yau that admits
an elliptic fibration.  If one obtains a light anti-self-dual
non-critical string in six dimensions, then in the reduction to five 
dimensions, the ``electric''
modes in five dimensions would come by wrapping the string on $\S^1$,
while the ``magnetic'' states in five dimensions would be unwrapped
states of the same string.  In the next section we study non-critical
strings in $F$-theory.

\newsec{Non-Critical Strings In $F$-Theory}

\def\F{{\bf F}}
A natural class of supersymmetric models in six dimensions is obtained
by compactifying the $E_8\times E_8$ heterotic string on K3,
with $12+n$ instantons in one $E_8$ and $12-n$ in the second.
(There is of course no essential loss in limiting to $n\geq 0$.)
According to \morrisonvafa, this model is equivalent to 
$F$-theory on the Hirzebruch surface $\F_n$.  That surface
can be described roughly as the quotient $\C^4/\C^*\times \C^*$, 
where $\C^4$ has complex coordinates $x,y,u,v$, and $\C^*\times \C^*$
acts by
\eqn\dufo{(x,y,u,v)\to (\lambda x,\lambda y, \mu u ,\lambda^n \mu v) ,}
with $\lambda,\mu\in \C^*$.  For the properties that we want to see,
it is helpful to give a ``symplectic'' description  of $\F_n$,
for which we introduce the $D$-functions
\eqn\dfn{\eqalign{ D_1 & = |x|^2+|y|^2+n|v|^2-r_1\cr
                   D_2 & = |u|^2+|v|^2 -r_2.\cr}}
Then one defines the $\F_n$ as the   space of solutions of $D_1=D_2=0$,
divided by $U(1)\times U(1)$; the $U(1)\times U(1)$ action is
given in \dufo, with now $\lambda,\mu\in U(1)$.  This exhibits
$\F_n$ as a Kahler manifold whose Kahler class depends on the two
real parameters $r_1$ and $r_2$.

The Hirzebruch surface is fibered over $\P^1$  by the map that
simply ``forgets'' $u$ and $v$; the fibers (obtained by projectivizing
$u-v$ space) are again copies of $\P^1$.  The generic section of
the fibration $\F_n\to \P^1$  is given by the equation
\eqn\kfn{v=u g_n(x,y),}
where $g_n$ is an arbitrary homogeneous polynomial of degree $n$.
There is also an ``exceptional section'' $E$ given by
\eqn\hufn{u=0.}
It is a copy of $\P^1$, embedded in $\F_n$ with self-intersection
number $-n$.  

The heterotic string compactified on K3 has for $n\not= 0$ 
a rather mysterious strong coupling singularity \minasian.  According
to \morrisonvafa, this singularity is associated in $F$-theory
with the collapse of the exceptional section $E$.  Notice
that on $E$, that is at $u=0$, the equations $D_1=D_2=0$ reduce
to $|v|^2=r_2$ and $|x|^2+|y|^2=r_1-nr_2$.  Thus, the area
of $E$ vanishes at $r_1-nr_2=0$.
When this happens, an $F$-theory string obtained by wrapping
a Type IIB threebrane around $E$ will go to zero tension.
Thus, as has been suspected on more generic grounds \refs{\seibergwitten,
\newduff}, a tensionless string appears at the point of the strong
coupling singularity.  

To understand more, we should know something about the singularity
that  $\F_n$ develops when one reaches $ r_1-nr_2=0$, with 
$r_1$ and $r_2$ positive.  With this condition on the
parameters, the equations \dfn\ imply
\eqn\hfn{|u|^2={1\over n}\left(|x|^2+|y|^2\right).}
Thus, $|u|$ is completely determined in terms of $x,y$.  Moreover, the
$U(1)$ symmetry associated with the parameter $\mu$  
in \dufo\ can be uniquely fixed (except at $x=y=u=0$, where
our subsequent assertions can be seen to hold anyway)
 by asking 
for $u$ to be, say, positive.  Thus one can eliminate and forget 
about $u$ and $\mu$, and describe this degeneration $\tilde \F_n$
of $\F_n$ in terms of three variables $x,y,v$, with one equation
\eqn\jumbo{|x|^2+|y|^2+n|v|^2 = r_1}
and one $U(1)$ symmetry to divide by, namely
\eqn\humbo{(x,y,v)\to (\lambda x,\lambda y,\lambda^nv).}

$\tilde \F_n$ has (for $n>1$) a singularity at the point $P$ with
 $x=y=0$.  This singularity is
simply a $\Z_n$ orbifold singularity; near $  x=y=0$, $\tilde \F_n$
looks like the quotient of the $x-y$ plane by the $\Z_n$ symmetry
\eqn\lumbo{(x,y)\to (\zeta x,\zeta y),~~{\rm with}\,\,\,\zeta^n=1.}
In fact, $\tilde \F_n$ is simply the weighted projective
space $\P^2_{1,1,n}$, the subscripts being the weights.  
To see this, one simply goes back from the symplectic description
of $\tilde \F_n$ by \jumbo\ and \humbo\ to the complex description of the
same space,
in which one replaces \jumbo\ by a condition that $x,y,$ and $v$ are
not all zero and permits $\lambda$ in \humbo\ to range over $\C^*$.

It is now possible to make some interesting statements about the nature
of the string that arises at the strong coupling singularity, for
various values of $n$.

\bigskip\noindent{$n=1$}

$n=1$ is the unique case in which there is no orbifold singularity;
in fact, $\tilde \F_1$ is the ordinary projective space $\P^2$, all
weights being one.  The point $P$ is a smooth point in $\P^2$,
which is replaced by a two-sphere $E$ of self-intersection $-1$ in
going from $\P^2$ to $\F_1$.  The operation of replacing a smooth
point $P$ by a two-sphere $E$ of self-intersection $-1$ is known
as ``blowing-up $P$,'' and is possible for any smooth point on
a complex surface.  We have just recovered the standard fact that
$\F_1$ is equivalent to the surface made by blowing-up  
$\P^2$ at a point,
to give an ``exceptional curve'' $E$.

To understand the strong coupling singularity for $n=1$, we should
study the string made by wrapping a Type IIB threebrane over $E$.
Analysis of this string only
depends on the behavior of $\F_1$ in a neighborhood of $E$.  But this
behavior is universal -- locally one would get the same picture
after blowing up any smooth point on any complex surface.  Thus,
the strong coupling singularity for $n=1$ has nothing really to do
with   the details of $\F_1$, and just involves the behavior of $F$-theory
under blow-down of a two-sphere of self-intersection number $-1$ 
to make a smooth point.

One of the things one most wants to know about the strong coupling
singularities of the heterotic string is whether, upon adjusting
a tensor multiplet to reach the strong coupling singularity, one
can make a transition to a ``Higgs banch'' with a different number
$n_T$ of tensor multiplets.  In \seibergwitten, it was shown using
anomalies that this could possibly occur only for $n=1$ and $n=4$.
(In this paper, the question will be addressed
only for $n=1$, but I understand that the $n=4$ case will be discussed
elsewhere \ref\morva{D. Morrison and C. Vafa, to appear.}.) 
As background, let 
us recall \vafa\ that in $F$-theory on a complex surface $B$,
 $n_T=b_2(B)-1$.  For instance,
the Hirzebruch surfaces have $b_2=2$ (corresponding to the two
Kahler parameters $r_1$ and $r_2$ introduced above), so $n_T=1$,
as expected for a perturbative heterotic string.  Under blow-up of
a point, $b_2$ increases by one, while under blow-down, $b_2$ decreases
by one.  

In particular $\P^2$ has $b_2=1$, so $F$-theory on $\P^2$ would
have $n_T=0$, no tensor multiplets at all.  This strongly hints
that the Higgs branch for $n=1$ is simply $F$-theory on $\P^2$.
We give some checks on this below, but first we point out some
general consequences of assuming that the Higgs branch exists
in this situation.  Since the whole analysis is local, if the transition
between $\P^2$ and $\F_1$ is possible in $F$-theory, analogous
transitions are possible for arbitrary blow-ups and blow-downs
(involving smooth points).  

For instance, one could start with $\F_n$ for any $n$, and blow up a
point $Q\in \F_n$, to get a surface $S$ with $b_2=3$,
corresponding to a model with $n_T=2$.  Then, if one
finds a two-sphere $J\subset S$ with self-intersection number $-1$, 
one can blow down $J$ to get another but perhaps different model
with $b_2=2$ and $n_T=1$.  For instance, the fibers of $\F_1\to \P^1$
are two-spheres with self-intersection number zero and so cannot
be blown down; but the fiber
$J$ containing $Q$ acquires self-intersection number $-1$ when $Q$
is blown up.  Thus, after blowing up $Q$, one can blow down $J$, getting
back to $n_T=1$.  In fact, the result of blowing up $Q$ and then
blowing down $J$ is to make a transition from $\F_n$ to $\F_{n-1}$
if $Q$ is generic, or to $\F_{n+1}$ if $Q$ lies in $E$. 

This should be compared to the situation seen in $M$-theory on
${\rm K3}\times \S^1/\Z_2$, with instanton numbers
$(12+n,12-n)$ at the two ends. Of  course,  this model is
 also believed to be
equivalent to the heterotic string on K3 and thus to $F$-theory
on $\F_n$.  In $M$-theory, one can apparently bring about a change
in $n$ by a process in which a small instanton   is emitted
from one of the ends, turning into a five-brane which then travels
to the other end of $\S^1/\Z_2$ and is reabsorbed.  Notice that
this is a two-step process and that the intermediate stage has $n_T=2$
(with one tensor multiplet carried by the five-brane), just
as in the $F$-theory process for changing $n$.  It is very natural
to suspect that these processes coincide, and thus, that the 
transition to and from the Higgs branch for $n=1$ is the same
as the $M$-theory transition involving a small $E_8$ instanton
that is emitted from the boundary.  

We will give further evidence for this below, but first we must
finally look a little more microscopically at $F$-theory in a neighborhood
of the exceptional curve $E$.
According to \morrisonvafa, to do $F$-theory on $\F_1$, we introduce
two more variables $X,Y$, transforming as
\eqn\polly{(X,Y)\to (\lambda^6\mu^4X,\lambda^9\mu^6Y),}
and write an equation
\eqn\olly{Y^2=X^3+f(x,y,u,v)X+g(x,y,u,v),}
where $f$ is of degree $(12,8)$ in $\lambda$ and $\mu$, and $g$ is
of degree  $(18,12)$.

Let us compare this to $F$-theory on $\P^2$.  For this, we use
homogeneous
coordinates $x,y,v$, scaling by $(x,y,v)\to (\lambda x,\lambda y,\lambda v)$,
and introduce two more variables $X,Y$ scaling
as $(X,Y)\to (\lambda^6X,\lambda^9Y)$, and we write an equation
\eqn\jolly{Y^2=X^3+\tilde f(x,y,v) X +\tilde g(x,y,v)}
where $\tilde f$ is of degree 12 and $\tilde g$ is of degree 18.

For instance, a typical monomial in $f$ is $x^{n_x}y^{n_y}u^{n_u}v^{n_v}$
with 
\eqn\hilly{\eqalign{ n_x+n_y + n_v & = 12 \cr
                       n_u+n_v& = 8.   \cr}}
By contrast, a typical monomial in $\tilde f$ is $x^{\tilde n_x}
y^{\tilde n_y}v^{\tilde n_v}$ with
\eqn\uhilly{\tilde n_x + \tilde n_y +\tilde n_v = 12.}
We see that for every possible monomial in $f$, there is a corresponding
possible monomial in $\tilde f$, with $\tilde n_x= n_x$, $\tilde n_y = n_y$,
and $\tilde n_v=n_v$.  But some monomials in $\tilde f$ are not associated
to any possible monomials in $f$ -- the missing ones are those with
$\tilde n_v>8$ (they would correspond to monomials in $f$ with $n_u<0$).
A simple count shows that there are $4+3+2+1=10$ monomials present
in $\tilde f$ and not in $f$.  Similarly, there are $6+5+\dots +1 = 21$
monomials present in $\tilde g$ and not in $g$.  Altogether, in going
from $\F_1$ to $\P^2$, we gain $10+21=31$ monomials.  

On the other hand, some coefficients in $f,g,\tilde f$, and $\tilde g$
can be eliminated by reparametrizations of the variables.  In $\P^2$,
one has a nine-dimensional group $GL(3)$ of linear transformations
of $x,y,v$.  For $\F_1$, the corresponding counting is a little trickier. 
There is a $GL(2)$ that acts on $x,y$ (four dimensional); one
also has $\delta u=\epsilon u$, $\delta v=\epsilon'v + \epsilon'' ux
+\epsilon''' uy$, making 8 reparametrizations in all; however, one
combination of these is a symmetry of $f$ and $g$, so only 7 parameters
can be removed by redefinition of $x,y,u$, and $v$.  

Thus, $F$-theory on $\P^2$ has 31 more monomials than $F$-theory on
$\F_1$, but two ($=9-7$) more of them can be removed by redefinitions of the
variables; so altogether the moduli  space of $F$-theory on $\P^2$
has $31-2=29$ more hypermultiplets than that of $F$-theory on 
$\F_1$.  The number 29 is the expected amount by which the number
of hypermultiplets must increase (to cancel anomalies) in a transition
in which the number of tensor multiplets decreases by one.  This
counting thus lends support to the idea that a transition is possible
from the ``Coulomb branch,'' $F$-theory on $\F_1$, to a ``Higgs branch,''
$F$-theory on $\P^2$.

Now let us try to check the idea that the phase transition in which
a point is blown up (passing for instance from $\P^2$ to $\F_1$) is the
same as the transition in which a small $E_8$ instanton is emitted
from the boundary in $M$-theory.  The light string in that transition
carries a rank eight current algebra, as explained in \hanany.
Let us look for such a current algebra in the string obtained in $F$-theory
by wrapping a Type IIB three-brane over a two-sphere $E$ of self-intersection
number $-1$.

$E$ is pierced by a certain number of Type IIB seven-branes.  The
current algebra will arise from intersections of Type IIB seven-branes
and three-branes.  So the first task is to count how many seven-branes
intersect $E$.  As in \vafa, we do this by counting parameters;
we ask how many parameters there are in the equation $Y^2=X^3+fX+g$
when restricted to $E$, and interpret these parameters as the positions
of the seven-branes.  Since we now only want to look at the structure
on $E$, we set $u=0$, so we only care about the monomials with $n_u=0$.
According to \hilly\ the  monomials in $f$ with $n_u=0$ have
$n_v=8$, $n_x+n_y=4$, giving five monomials in all; likewise there
are seven relevant monomials in $g$, and so $5+7=12$ in all.
After removing four parameters that can be absorbed in $GL(2)$
transformations of $x,y$, we have $12-4=8$ relevant parameters
that we interpret as positions of seven-branes, so there are eight
such seven-branes.  

What remains is a perturbative string computation 
(in Minkowski space with coordinates $x^0,\dots,x^9$)
involving the transverse intersection of a Type IIB three-brane
at, say, $x^4=\dots = x^9=0$ with a Type IIB seven-brane at, say,
$x^2=x^3=0$.  Obviously, their intersection is the cosmic string
given by $x^2=\dots = x^9=0$. 
(We call it a cosmic string to avoid confusion with the elementary
Type IIB strings that will enter momentarily.)
In working out the excitation spectrum
along this cosmic string, one must quantize elementary Type IIB strings
that start on the three-brane and end on the seven-brane, or vice-versa.
For either of these orientations, a standard free field calculation
shows that there are no massless bosons, and a single massless fermion
that is left-moving along the cosmic string.  
(The calculation is actually isomorphic to the analysis of the DN
sector for Type I Dirichlet one-branes; see the concluding paragraphs of
\ref\polch{J. Polchinski and E. Witten, ``Evidence For Heterotic
- Type I String Duality,'' hepth/9511157.}.)
Allowing for both
orientations, one gets two left-moving fermions, that is a rank
one current algebra, on the intersection of a seven-brane with a three-brane.

In our $F$-theory problem, the non-critical string comes from a three-brane
that has, as we saw above, transverse intersections with {\it eight}
seven-branes, so it carries altogether a rank eight current algebra.  This
gives strong support to the idea that this string is the same
as the one that is associated to small $E_8$ instantons.

\bigskip\noindent{$n=2$}

Now we move on to $n=2$.  
The main difference is that the exceptional curve $E$ now
has self-intersection number $-2$, so that blowing it down
gives a $\Z_2$ orbifold singularity, the quotient of the $x-y$ plane
by $(x,y)\to (-x,-y)$.  

The crucial property of this singularity is that it looks locally
like a singularity of a hyper-Kahler manifold,
simply because the holomorphic two-form $dx\wedge dy$ is invariant
under the $\Z_2$.  (This property will not recur for any $n>2$.)
This particular singularity is called an $A_1$ singularity.  

Now, recall that $F$-theory is simply Type IIB superstring theory
with a variable complex coupling ``constant.''  The coupling varies in such
a way as to compensate for the curvature of the space-time.  If one
takes the ten-dimensional space-time to be $\R^6\times B$, with
$B$ a hyper-Kahler manifold, then there is no reason for the coupling
 to vary on $B$, so in this situation $F$-theory 
reduces to ordinary Type IIB theory.  

$\F_2$ is not hyper-Kahler, so $F$-theory on $\F_2$ is not equivalent
(in any evident way) to an ordinary Type IIB theory. However,
in a neighborhood of $E$, $\F_2$ does look like a hyper-Kahler manifold.
In particular, in $F$-theory on $E$, the Type IIB coupling 
is {\it constant} in a neighborhood of $E$.  This constant is arbitrary,
and can be taken to be small.  Thus, the strong coupling singularity
for $n=2$ is equivalent to a phenomenon that can occur in weakly
coupled Type IIB superstring theory.  It is simply the behavior of
the Type IIB theory as one approaches an $A_1$ singularity.  

The strong coupling singularity of the heterotic string for $n=2$,
that is, for instanton numbers $(14,10)$, thus involves the appearance
of the same non-critical string that has been studied for Type IIB
at an $A_1$ singularity \refs{\ewitten,\otherstrom}.  Because
Type IIB has twice as much supersymmetry as the heterotic string,
this string can carry $(0,2)$ spacetime supersymmetry in six dimensions,
even though we are finding it in an $F$-theory model that has only $(0,1)$
spacetime  supersymmetry.  Some consequences of the extra supersymmetry
were discussed in \seibergwitten.  This string is controlled by
five relevant parameters (the scalars in a $(0,2)$ tensor multiplet)
rather than one for the non-critical strings with $n\not= 1$.
As a result, one can ``go around'' the singularity, unlike the other
cases.  The extra parameters correspond to the non-polynomial deformation
of $F$-theory on $\F_2$ which was used in \morrisonvafa\ to show
that the $n=2$ and $n=0$ models are the same.  At the critical
point of this string, there is a $\Z_2$ symmetry, familiar
in the Type IIB description, which in the present context is the
strong-weak  coupling symmetry of the heterotic string for
instanton numbers $(12,12)$ or $(14,10)$.  In the $(14,10)$ case,
the existence of this symmetry was first suggested in \ref\font{
G. Aldazabal, A. Font, L. E. Ibanez, and F. Quevedo,
``Heterotic/Heterotic Duality In $D=6, \,D=4$,'' hepth/9602097.}.

\bigskip\noindent{$n>2$}

Now we consider the case of $n>2$.  The new ingredient is that $\tilde \F_n$
does not look like a Calabi-Yau manifold near its singularity.
The singularity of $\tilde \F_n$ 
is now the orbifold singularity obtained
by dividing the $x-y$ plane -- which we will call $W$ -- by
$(x,y)\to (\zeta x,\zeta y)$, with $\zeta^n=1$.
The holomorphic two-form $dx\wedge dy$ on $W$ is multiplied  
by $\zeta^{2n}$ under this operation.

This is the situation where, in $F$-theory, one restores the
Calabi-Yau property by letting the coupling ``constant'' vary with position.
Instead of $\tilde \F_n$, one considers a Calabi-Yau 
manifold $Z$ that maps to 
$\tilde \F_n$, with the generic
fibers being two-tori, whose complex structure
is determined by the expectation value of the scalars of the Type IIB
theory. In our case, we only need to know the local behavior near
the singularity, so it is enough to find a Calabi-Yau manifold
fibered by two-tori over $W/\Z_n$.

The most obvious thing to do is to take a {\it constant} two-torus
$A$, with $\Z_n$ action, and look at $(W\times A)/\Z_n$, which
maps to $W/\Z_n$ (by the map that forgets $A$) with the generic fiber
being a two-torus, that is, a copy of $A$.  For $(W\times A)/\Z_n$ to
be a Calabi-Yau manifold, the $\Z_n$ action on $A$ should be such
that a holomorphic differential $\lambda$ transforms under
$(x,y)\to (\zeta x,\zeta y)$ as $\lambda\to \zeta^{-2}\lambda$.
This is possible if and only if $\zeta^2$ is of order $2,3,4$, or 6.
Thus, $n$ must be 3,4,6,8, or 12.  These are the values of $n$ studied
in \morrisonvafa, and the only ones that will be considered here.
\foot{The other possible 
values of $n$ for  $E_8\times E_8$ heterotic
strings on K3 with instanton numbers $(12+n,12-n)$  are 5 and 7;
I understand that results have been obtained for those values of $n$
\morva.}

The conclusion, then, is that for $n=3,4,6,8,$ or 12, what happens
at the strong coupling point is the appearance in space-time of what
at least macroscopically looks like an
 orbifold singularity.
  In constructing this orbifold,
one must use the $SL(2,\Z)$ symmetry of the Type IIB theory.
One finds this singularity, at least macroscopically,
 by dividing the theory by an operation
that acts on the $x-y$ plane by $(x,y)\to (\zeta x,\zeta y)$ 
together with an $SL(2,\Z)$ transformation which, for $n=3,4,6,8,$ or 12,
is of order $a_n=3,2,3,4$, or 6.  

To be more precise about what one means here by an ``orbifold,''
recall how orbifolds enter in perturbative string theory:
one can make a non-singular orbifold conformal field theory; or
by shifting a theta angle (as in \ref\aspin{P. Aspinwall, ``Enhanced
Gauge Symmetries And K3 Surfaces,'' hepth/9507012.}),
one can get   a conformal field theory singularity where non-perturbative
states may become massless. 
Here our ``orbifold'' (whose theta angle is frozen at the critical
value) corresponds to a truly singular configuration, since it
 supports a tensionless string.  In any event, the two types
of orbifold differ only in the structure near the space-time singularity,
so macroscopically when the heterotic string gets a strong coupling 
singularity the $F$-theory description really does develop  an orbifold
singularity.

We can now see that the case $n=4$ is exceptional.  Only special 
two-tori have symmetries that act with order 3, 4, or 6 on a holomorphic 
differential; only for unique values of the coupling constants does
the Type IIB theory have such symmetries.  Thus, for
$n=3,6,8$, or 12, the behavior near the strong coupling singularity
of the heterotic string involves the Type IIB theory at a special
strong coupling point, one of the orbifold points in the ten-dimensional
Type IIB moduli space.  It may be that at one of those special strong
coupling points, the Type IIB theory has some unusual dynamics.
No such dynamics is possible in conventional field theory, since the
ten-dimensional supersymmetry does not permit any extra particles to
become massless at a special value of the coupling, but it is conceivable
that something novel 
happens in string theory.  If that is the case, one may need
to first understand this behavior to get
a real understanding of the heterotic string strong coupling singularity
for $n=3,6,8$, or 12.

On the other hand, for $n=4$, $a_n=2$, and since a generic two-torus
has a symmetry (``multiplication by $-1$'') that acts with order two
on a holomorphic differential, the effective Type IIB coupling can
have an arbitrary value in the region of space that is important
for understanding the heterotic string strong coupling singularity.
It should thus be possible to come relatively close to understanding 
in weak coupling
the non-critical string relevant to $n=4$.  In fact, the element
$-1$ of $SL(2,\Z)$ which we need to use acts as $-1$ on, for instance,
the Neveu-Schwarz two-form of the theory; the symmetry that does
this is the reversal of world-sheet orientation,
the exchange of the left- and right-movers of the theory.
The singular space-time at the $n=4$ strong coupling singularity
is thus at least macroscopically a kind of orientifold 
made by dividing  the $(x,y)$ plane
by $(x,y)\to (ix,iy)$ while also reversing the Type IIB world-sheet
orientation.  The relevant operation reversing the 
world-sheet orientation is actually of order four in acting on the fermions.

Note that the $n=3$ example involves a sort of $\Z_3$ orbifold closely
related to an $M$-theory example studied in section 2.4.

\newsec{Some Examples In Four Dimensions}

This paper has been mainly concerned with phase transitions in
five and six dimensions, but I will here briefly point out that
critical points with tensionless strings will also be common in four 
dimensional models with $N=1$ and $N=2$ spacetime supersymmetry.

\def\O{{\cal O}}
For Calabi-Yau compactification of
Type II, the most obvious example is to consider the Type IIB
theory at a point where the Type IIA theory gets a massless charged
hypermultiplet, or vice-versa.  It is then obvious by an argument
as in \ewitten, involving the equivalence of the two theories after
further compactification on a circle, that Type IIB gets a tensionless
string at such a point.  More concretely, Type IIA gets a massless
hypermultiplet by wrapping a two-brane around a collapsing two-cycle,
while Type IIB gets a tensionless string by wrapping a three-brane
over the same collapsing two-cycle.  

An important example of a collapsing two-cycle is the one that arises
at a conifold singularity reached by varying Kahler parameters.
The collapsing two-cycle is then a two-sphere of normal bundle 
$\O(-1)\oplus \O(-1)$.  (This is the ``generic'' normal bundle for
a holomorphic two-sphere in a Calabi-Yau three-fold.) 
The string that appears here of course carries $N=2$ supersymmetry,
since we have found it in Calabi-Yau compactification of Type II.
It is a fairly close
analog of the tensionless string that one gets in Type IIB on K3
when a two-cycle collapses.

Now we want to get the {\it same} string in a Calabi-Yau compactification
of the heterotic string.  (Thus, we will be finding a string
that carries $N=2$ supersymmetry in an $N=1$ model, as happened
for $n=2$ in section 3.  Many variations that will not be explored here
give strings that only carry $N=1$ supersymmetry.)  
First of all, to apply Type IIB techniques
to the heterotic string, we will, of course, use $F$-theory
as in \refs{\vafa,\morrisonvafa}.  We need a complex threefold $W$
with the following properties:

(1) $W$ has sufficiently positive curvature that one can do $F$-theory
with  $W$ as base; there should be a Calabi-Yau fourfold that maps to $W$ with
generic fiber a two-torus.

(2) There should be a map $W\to B$ with $B$ a complex surface and the
generic fiber being a $\P^1$.  In that case, one can fiber-wise
use the relation \vafa\ between Type IIB on $\P^1$ (with seven-branes
and a variable coupling) and the heterotic string on $\T^2$.  
(In our example, the fibers of $W\to B$ will {\it all} be two-tori,
so this fiber-wise transformation is justified simply by an adiabatic
argument.) 
Thus, on
replacing $W$ with a Calabi-Yau  three-fold $Z$ that maps to $W$ with
generic fiber a two-torus, Type IIB on $W$ will be equivalent to the
heterotic string on $Z$.

(3) $W$ will have a holomorphic two-sphere $E$, with normal bundle
${\cal O}(-1)\oplus {\cal O}(-1)$, 
which collapses as a Kahler parameter is varied.

Condition (3) puts us in the situation we encountered in section 3 for
$n=2$.  Though $W$ is not a Calabi-Yau manifold, a neighborhood of $E$
looks like one, so near $E$ the Type IIB coupling is {\it constant},
and one can study the collapse of $E$ using weakly coupled Type IIB
string theory.  Therefore, one will get the same tensionless string
as in collapse of a two-sphere with local structure of $E$ in Calabi-Yau
compactification of Type IIB.

To obey condition (1), we start with $\C^6$ with coordinates $x_1,\dots ,x_6$
and the scalings
\eqn\pully{(x_1,x_2,x_3,x_4,x_5,x_6)\to (\lambda x_1,\lambda x_2,
\mu x_3,\mu x_4,\nu x_5,\nu\lambda\mu x_6).}
Either one divides by these scalings with
 $\lambda,\mu,\nu\in (\C^*)^3$ and omits certain
linear subspaces of $\C^6$ or -- more helpful for exhibiting the Kahler
parameters -- one takes $\lambda,\mu,\nu\in U(1)^3$ and imposes
the $D$-field equations:
\eqn\kully{\eqalign{ |x_1|^2+|x_2|^2+|x_6|^2 & = r_1  \cr
                     |x_3|^2+|x_4|^2 +|x_6|^2 & = r_2 \cr
                      |x_5|^2+|x_6|^2 & = r_3.\cr}}
Our space $W$ is just the space of solutions of \kully\ (with suitable
$r_i$) divided by $U(1)^3$.
To do $F$-theory with $W$ as base, we introduce two new variables $X,Y$
and an equation
\eqn\jully{Y^2=X^3+f(x_1,\dots,x_6)X +g(x_1,\dots,x_6),}
where (to get a Calabi-Yau four-fold) $f$ is of degree $(6,6,4)$ and
$y$ of degree $(9,9,6)$ in $\lambda,\mu,\nu$.  I will leave it to the
reader to verify that generic such $f$ and $g$ give a smooth Calabi-Yau
four-fold corresponding to a model that generically has the gauge
symmetry completely broken.

\nref\kachru{S. Kachru and C. Vafa, ``Exact Results For $N=2$
Compactifications Of Heterotic Strings,'' Nucl. Phys. {\bf B450}
(1995) 69.}
\nref\multiplefiber{
P. S. Aspinwall and M. Gross, ``Heterotic-Heterotic String
Duality And Multiplet K3 Fibrations,'' hepth/9602118.}
To verify condition (2), note that forgetting $x_5,x_6$ gives
a map from $W$ to $B=\P^1\times \P^1$, the fibers being (all)
$\P^1$'s obtained by projectivizing $x_5,x_6$.  Thus, Type IIB
on $W$ is equivalent to the heterotic string on a Calabi-Yau $Z$
that is elliptically fibered over $\P^1\times \P^1$.
This particular Calabi-Yau has played a prominent role in the last
year \refs{\kachru,\morrisonvafa,\multiplefiber}.

To verify condition (3), we take $r_2>0$, $r_3>r_2$, $r_1>r_2$,
and let $E$ be the two-sphere defined by $x_3=x_4=0$.  We can uniquely
solve the last two equations in \kully\ and fix the $U(1)^2$
associated with $\mu$ and $\nu$ by taking $x_6=\sqrt{r_2},$
$x_5=\sqrt{r_3-r_2}$.  Then $x_1$ and $x_2$, modulo scaling by $\lambda$,
parametrize a two-sphere $E$ of Kahler class proportional to  $r_1-r_2$.
$E$ can be seen to have normal bundle $\O(-1)\oplus \O(-1)$.
Now note that in
 the limit $r_1\to r_2$ (keeping $r_3>r_2>0$), $E$ collapses.
In fact, if we continue to $r_1<r_2$, $E$ disappears and is replaced
by a two-sphere $F$ defined by $x_1=x_2=0$; $F$ is
 also of normal bundle $\O(-1)\oplus \O(-1)$,
with Kahler class proportional to $r_2-r_1$.  This situation is
a typical Type II ``flop'' (note the change in sign of the area). 
At $r_1=r_2$, Type IIB on $W$ and therefore
also the heterotic string on $Z$ has the  tensionless string
discussed at the beginning of this section.
 
\bigskip
I benefited from numerous discussions with N. Seiberg. I would
like also to thank D. Freed and 
D. Morrison for helpful explanations
of several points.
\listrefs
   \end